\documentclass[aps,showpacs,preprintnumbers,amsmath,amssymb, 
eqsecnum,
twocolumn, 
tightenlines,
]{revtex4}

\usepackage[dvips]{graphicx}
\usepackage{bm}

\begin{document}

\bibliographystyle{apsrev} 

\title {Two heavy fermions bound 
via Higgs boson exchange}

\author{V. V. Flambaum} \email[Email:]{flambaum@phys.unsw.edu.au}
\author{M. Yu. Kuchiev} \email[Email:]{kmy@phys.unsw.edu.au}

\affiliation{School of Physics, University of New South Wales, Sydney
  2052, Australia}

    \date{\today}

    \begin{abstract}
A system of two heavy fermions, leptons or quarks of the fourth generation, which are bound together via the Higgs boson exchange is studied. The conventional Yukawa-type interaction produced by this exchange is accompanied by several important corrections. We derived the Hamiltonian, which describes the correction arising from the retardation (compare the Breit correction in QED); we also calculated the relativistic and radiative corrections. The Higgs-induced bound state appears for the fermion mass  $m>m_\text{cr} \approx 500$ GeV. When the long-range Coulomb interaction or the gluon exchange are included, the bound states exist for any mass, but the Higgs exchange drastically increases the binding energy of these states when $m$ is approaching $m_\text{cr}$. In the region $m>m_\text{cr}$ the gluon exchange gives a sizable correction to the Higgs induced binding energy. This correction greatly exceeds typical binding energies in the states produced via the gluon exchange only. 
The final results for the binding energies are presented as analytical functions of the Higgs and fermion masses.
 The possibility of detection of the considered bound states at LHC is discussed. 
       \end{abstract}


%

    \maketitle

  \section{introduction}
   \label{intro}
Heavy  fermions with masses 300-500 GeV and even heavier have been suggested recently to explain new results on CP violation in B decays \cite{B}. 
The bags with heavy fermions and bosons may play a crucial role in the bariogenesis \cite{Kuchiev:2008gt,Flambaum:2010}. The nature of bound states made of such  heavy particles is different from that for usual hadrons. 
The strength of the attraction between heavy particles due to the Higgs boson  exchange increases with the particle mass. This may lead to formation of a new type of bound states.
The Higgs-induced bags made of heavy fermions and bosons have been considered in numerous publications - see e.g. 
\cite{
Vinciarelli:1972zp,%
PhysRevD.9.2291,%
Chodos:1974je,%
Creutz:1974bw,%
Bardeen:1974wr,%
Giles:1975gy,%
Huang:1975ih,%
PhysRevD.15.1694,%
PhysRevD.25.1951,%
PhysRevLett.53.2203,%
PhysRevD.32.1816,%
Khlebnikov:1986ky,%
Anderson:1990kb,%
MacKenzie:1991xg,%
Macpherson:1993rf,%
Johnson:1986xz,%
Froggatt:etal,%
Kuchiev:2008fd,%
Richard:2008uq,%
Crichigno,%
Kuchiev:2008gt,%
Crichigno:2010ky,%
Kuchiev-1012.0903,%
Kuchiev:2011,%
critical-12-bag%
}. 
An alternative, but relative line of recent research addressed the 
problem of a possible dynamical electroweak symmetry breaking that may take
place if the fourth generation is present \cite{Hung:2010xh}.

  In 
several 
our previous works we concentrated mainly on ``magic"  
bag consisting of 6 heavy  quarks and 6 antiquarks (altogether $N=3\times2\times2=12$ particles) occupying $1s$ shell. 
Its advantage is that it gives arguably the lowest 
limit on the fermion mass, which guarantees the binding.
Besides, such multifermion bags may play a role in bariogenesis.
However, these fermion bags are very difficult targets for possible 
observation at accelerators. 
Having this in mind in the present paper we 
consider the Higgs-induced bound states of only two
fermions.
The problem has recently been addressed in Ref. \cite{Ishiwata:2011ny}, where
the relativistic as well as radiative corrections induced by the 
fermion polarization were considered.
To make our calculations accurate and reliable we 
consider a number of important corrections including
the relativistic and retardation corrections,
as well as the radiative corrections.
It should be emphasized that the retardation plays a very essential role in the problem,
which makes it necessary to consider it alongside the relativistic correction.
To describe the joint contribution of the relativistic and retardation corrections 
we derive the nonrelativistic Hamiltonian for the Higgs-induced potential,
which includes all  $v^2/c^2$ relativistic corrections. 
This approach is similar in nature to the way the Breit
correction is usually accounted for in QED. 
Our Hamiltonian-based treatment provides the convenient flexibility
which allows it to be applied to a variety of similar problems in the future.

To calculate the radiative corrections we 
employ the conventional methods developed previously 
to calculate the Lamb shift in atoms,
see e.g. \cite{QED}. 
Following the calculations of \cite{Kuchiev-1012.0903} 
we express the radiative correction in terms
of the renormalized Higgs vertex correction, which gives the fermion self-energy,
and the renormalized Higgs polarization operator for the fermion loops.  
It should be stressed that the vertex correction derived for the first time in \cite{Kuchiev-1012.0903} proves to be very large and has the  sign opposite to the polarization correction, which makes it necessary to include it in the calculations. 
(Compare the Lamb shift problem, where the self-energy correction dominates).
We solve the Schr\"{o}dinger  equation with an effective Hamiltonian  which includes the nonrelativistic Yukawa Hamiltonian, the Breit-type  relativistic terms and the radiative corrections, and find the ground state.

Due to the short-range character of the interaction, the Higgs-induced
bound state is formed only if the particle mass exceeds the limit, which is found to be
$m>m_\text{cr} \approx  ( m_h /100 \,\mathrm{GeV})^{1/3} \,500 \,\mathrm{GeV}$.
When the long-range Coulomb interaction or the gluon exchange are included,
the bound states exists for an arbitrary mass.
We found that the Higgs-induced interaction produces an
an interesting effect on these states even for $m < m_\text{cr}$. The Higgs 
induced interaction dramatically increases the binding energy and reduces the size of
 the bound state. 
For $m =m_\text{cr}$ the binding energy increases
$\alpha_h/\alpha$ times and the size decreases $\alpha_h/\alpha$ times where
 $\alpha_h$ is the constant of the Higgs exchange potential
and $\alpha$ is the constant of the long-range potential.
  
   \section{non-relativistic approximation}
   \label{nonrel}
   
\subsection{Yukawa potential}
  
  It is convenient to start from the 
simple non-relativistic approximation. 
In the center of mass reference frame the interaction between two fermions due to the Higgs boson exchange may be described by the following Hamiltonian ($\hbar=c=1$)
\begin{equation}
\label{Hh}
H_0= \frac{{\bf p}^2}{ m } + V_\text{Y}(r)
\end{equation}
where $V_\text{Y}(r)$ is the Yukawa-type interaction induced by the Higgs exchange
\begin{equation}
V_\text{Y}(r) \,=\, - \frac{\alpha_h }{r}\,\exp{(-m_h \,r)}~.
	\label{Vh}
\end{equation}
Here $m_h$ is the Higgs mass, and  $\alpha_h=g_h^2/(4\pi)$ is the effective constant describing the strength of the interaction. Within the Standard model  $g_h=m/v$
where $v=246$ GeV is the Higgs VEV. 

Numerical solution of the Schr\"{o}dinger   equation with the Hamiltonian (\ref{Hh}) 
gives the bound state for  $\alpha_h m/2m_h>0.84$. This corresponds to the particle mass
\begin{equation}
\label{mcr}
 m>m_\text{cr}\,=\,2 v (\,2.64 \,m_h v^2)^{1/3}\approx 2.76 \,(\,m_h v^2)^{1/3}~.
 \end{equation}
A similar estimate was made previously in \cite{Hung:2010xh}.
 For $m_h=100$ GeV  we obtain $m_\text{cr}\approx  503$ GeV. 
This is larger than the critical mass $m_\text{cr, bag}=320$ GeV, which was found in \cite{critical-12-bag} for the bag of 12 fermions, but the distinction is not dramatic.
To put it in perspective 
it is instructive to compare it with 
the situation when the bound state of the pair is produced by
the heavy fermion interacting with some known, relatively light  particle,
when the critical mass proves to be much larger.
For example, when the heavy fermion interacts with the t-quark 
the critical mass of the heavy particle is over 2 TeV. In the present work we do not consider so heavy particles.

\subsection{Adding Coulomb and gluon potentials} 

The Higgs potential may be  important even for  $m<m_\text{cr}$. Indeed, the bound state for $m<m_\text{cr}$ may be produced by a relatively weak long-range interaction 
\begin{equation}
V_\text{lr}(r) \,=\, - \frac{a }{r}~,
	\label{Vl}
\end{equation}
where $a=\alpha$ 
for the electrostatic interaction between charged leptons  or
\begin{equation}
a\,=\,\frac{4}{3}~\alpha_s  
\label{Vg}
\end{equation}
 due to the gluon-exchange 
 \cite{sigma}.
We restrict our attention here by the color singlet state of the pair. 
Generally speaking the pair of quarks can be in the octet state as well. 
However, it is known that the effective potential for the octet is repulsive, and what is worse, the octet pair of heavy quarks cannot exist by itself; its colored state needs the presence of additional, probably light quarks, which would compensate it.
 
The constant $a$ in Eqs.(\ref{Vl}), (\ref{Vg}) is
substantially smaller than  $\alpha_h(m_\text{cr})=0.33$. 
It is shown below that when the particle mass $m$ is approaching  $m_\text{cr}$, the binding energy increases  $\alpha_h/a$ times.
 
To estimate the energy
we start from the  variational approach, which gives accuracy about 10\% (below we will perform accurate numerical calculations). 
Take the variational ground state wave function in the hydrogen-like form 
\begin{equation}
\psi(r)= \pi^{-1/2} q^{3/2}  \exp{(-q \,r)}~.
\label{psi}
\end{equation}
Then from the Hamiltonian (\ref{Hh}) one finds the following expectation value for the energy 
 \begin{equation}
 \label{Eh}
E_0(q)= \frac{q^2}{ m } - \frac{4 \alpha_h q^3}{(m_h+2q)^2}
\end{equation}
The minimization of $E_0(q)$ gives the equation on $q$
  \begin{equation}\label{qh}
\frac{	\partial E_0(q)}{\partial q}= \frac{2q}{ m } - \frac{12 \alpha_h q^2}{(m_h+2q)^2} +\frac{16 \alpha_h q^3}{(m_h+2q)^3}=0
\end{equation}
Let us first find the critical condition for the bound state to appear. Combining equation  (\ref{qh}) with $E_0(q)=0$ one finds 
\begin{align}
&q_\text{cr}=m_h/2,
\label{mh/2}
\\
& \big(\alpha_h \,m\big)_\text{cr}/2 = m_h
\label{mh}
\end{align}
A similar estimate was made previously in \cite{Hung:2010xh}. The last condition allows one to find for the critical mass 
$m_\text{cr}=2(\pi m_h v^2)^{1/3}\approx 2.93\, (m_h v^2)^{1/3}$, which only slightly, by
$6\%$ exceeds the corresponding value in (\ref{mcr}) extracted from numerical calculations.

Let us add to the energy the contribution of the  long-range potential $V_\text{lr}=-a/r$. 
Then we need to find the eigenvalue $E$ of the Hamiltonian
$H=H_0+V_\text{lr}$.
Following the variational approximation we find then
\begin{align}
E(q)=\, &\frac{q^2}{ m } - \frac{4 \alpha_h q^3}{(m_h+2q)^2} -a q~,
\label{Ea}
\\
\frac{	\partial E(q)}{\partial q}=\,& \frac{2q}{ m } - \frac{12 \alpha_h q^2}{(m_h+2k)^2} +\frac{16 \alpha_h q^3}{(m_h+2k)^3}-a =0
\label{qa}
\end{align}
In the absence of the short-range potential, when  $\alpha_h=0$, these equations obviously reproduce the exact Coulomb-type wave function with $q=m a/2$ and energy $E=E_C=-a^2 m/4$. 

An interesting phenomenon takes place in the opposite case, when $\alpha_h \gg a$. 
Let us choose
$q=q_\text{cr}=m_h/2$, where 
$q_\text{cr}$ is critical value, which was obtained in Eq.(\ref{mh/2})  for the problem
without the long range interaction.
With this $q$ the energy equals $E=-a q=-a m_h/2 =-\alpha_h a m/4$, where Eq.(\ref{mh}) was used. 
Thus the Higgs exchange increases the binding energy of the system by a factor $E/E_C=
a_h/a$. Correspondingly, the size of the system decreases by the same amount.
We conclude that the Higgs contribution proves to be very important even well before the Higgs exchange alone would lead to the binding. For leptons bound by the Coulomb attraction the Higgs exchange produces $\sim$ 30 times enhancement of the binding near the critical mass.

\subsection{Confining potential}
The variational approach may also be used to estimate contribution of other interaction terms.
In particular, in the region of large separation between the quarks 
the effect of the confinement could become essential.
Note though that the region of large separations produces only
small corrections to the energy.
Therefore with the sufficient accuracy we can approximate an influence of
the confinement using the simple approximation, by applying the confining potential in its most transparent linear form $V_\text{conf}=\sigma r$, where $\sigma=0.2$ GeV$^2$ \cite{sigma}. We find then that the contribution to the 
 energy equals  $\delta E_\text{corr}= 3\sigma/q$. For $q>q_\text{cr}=m_h/2$ this contribution is by three orders of magnitude smaller than the effect produced by the pure gluon exchange, which equals $(4/3)\alpha_s q $.

\subsection{$Z$-boson contribution} 

The weak interaction produces the short-range potential via the $Z$-boson exchange,  
which with the sufficient for our purposes accuracy can be approximated as follows
\begin{equation}
V_Z(r) \,\approx\, - \frac{\alpha_Z }{r}\,\exp{(-m_Z \,r)}~.
	\label{VZ}
\end{equation}
Deriving this expression the low-energy, nonrelativistic scattering of the pair was considered in the unitary gauge.
Using the variational approach one finds that (\ref{VZ}) gives 
the following contribution to the energy
\begin{equation}\label{EZ}
V_Z(q) \approx- \frac{4 \alpha_Z q^3}{(m_Z+2q)^2} 
\end{equation}
For $q>q_\text{cr}=m_h/2$ it is comparable to the 
contribution due to the photon exchange (which can be approximated by the electrostatic interaction) 
since the short-range suppression factor $(m_Z/2q+1)^{-2}$ is close to 1. 
Hence the effect produced  by the $Z$-boson exchange 
is much smaller than the one originating from the Higgs boson exchange. 
One of the consequences is that a bound state of heavy neutrinos (from the hypothetical fourth generation) may be formed only if $m>500$ GeV, and this bound state would 
be almost entirely due to the Higgs boson exchange.

\subsection{Higgs contribution for $m <m_\text{cr}/2$}

If mass of the particle is significantly smaller than $m_\text{cr}$ the system is dominated
by the long range interaction, i.e. $q\approx m a/2 \ll m_h/2$,  and  the short-range Higgs exchange
 is  suppressed by the factor $(\alpha_h/a)(ma/m_h)^2$. The Z-boson contribution has a similar suppression.  

\subsection{Binding energy for   $m>m_\text{cr}$}
 
A relatively accurate result in the area of  $m>m_\text{cr}$ may be obtained by using the expansion in 
powers of
$m_h/2q$. Indeed, near the minimum the energy $E(q)$ is not sensitive to minor variations of $q$, and it is sufficient to find an approximate value of $q$. The error produced in this approximation of $q$ may be smaller than the error of the variational approach itself. Keeping three leading terms  in the
  $m_h/(2q)$ expansion of the Higgs contribution in 
Eqs. (\ref{Ea}), (\ref{qa})  we derive an approximate value 
   \begin{equation}
q \,\approx \,\frac{m}{2}\,(\tilde \alpha_h +a) ~,
	\label{qapprox}
\end{equation}
 where ${\tilde \alpha_h}$ is the effective value of the interaction strength suppressed
 by the short-range character of the Higgs exchange, 
     \begin{equation}
{\tilde \alpha_h} =\alpha_h\left(1+ 3 \frac{m_h^2}{m^2(\alpha_h +a)^2}\right)^{-1}.
	\label{tildealpha}
\end{equation}
Substituting $q$ from Eq.(\ref{qapprox}) into (\ref{Ea}) we find
the energy, which for $m>0.75$ TeV differs from the results of direct numerical solution of the Schr\"{o}dinger equation 
only at the per cent level. 

\subsection{Binding energy in the limit $m \gg m_h$} 
In the limit 
 $m \gg m_h$ the non-relativistic energy tends to 
   \begin{equation}  
  E_0\approx -(\alpha_h+a)^2 m/4  
\label{Elarge}
\end{equation} 
However, the relativistic corrections ( $ \propto  \alpha_h ^2$) become large in this limit since $\alpha_h \propto  m^2$. For example, $\alpha_h=m^2/4 \pi v^2=1$ at $m=0.87$TeV. 
In the following sections we will improve our results to account for this fact.
 
 \section{relativistic corrections}
   \label{relativisticB}
 
 \subsection{Relativistic and retardation effects}
 \label{Relativistic-retardation}
\begin{figure}[tbh]
  \centering \includegraphics[ height=2.5cm, keepaspectratio=true, angle=0]{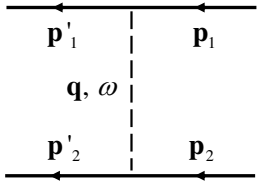}
\vspace{0cm}
\caption{Higgs boson exchange. The scattering process for two fermions is used in the text 
as a tool, which allows one to calculate the retardation and relativistic corrections. }\label{fig1} 
   \end{figure}

It is convenient to combine together the relativistic corrections 
for the propagation of fermions and the retardation correction for the 
propagation of the intermediate Higgs boson. We  rely on the 
conventional technique, similar to the one applied for
derivation of the Breit corrections in QED.  
Consider first the scattering process for the two fermions interacting through
the Higgs boson exchange, see the diagram in Fig. \ref{fig1}, derive for this process the necessary relativistic and retardation corrections, and  
after that reformulate the result to make it applicable to the
bound state problem.

   \noindent
The analytical expression for the amplitude of the scattering process
illustrated by Fig. \ref{fig1} can be written as the matrix element 
\begin{equation}
M\,=\,\langle u_1^\prime, u_2^\prime\,|\,V({\bf q})\,|\,u_1, u_2  \rangle~,
\label{M}
\end{equation} 
in which  $|u_i \rangle$ and $\langle u_i^\prime| $ 
are spinors that describe the incoming and outgoing fermions,  while
$V$ represents the Higgs 
propagating in the intermediate state
\begin{align}\label{Int1}
V({\bf q})\,=\, & g^2_h~\frac{ \beta_1 \beta_2}{\omega^2-{\bf q}^2 -m_h^2} 
\\
\nonumber
\,\approx\,&
-g^2_h \,\frac{\beta_1 \beta_2}{{\bf q}^2 +m_h^2}\,\left(1 + \frac{\omega^2}{{\bf q}^2 +m_h^2}\right)
\end{align} 
Here $\beta_i$, $i=1,2$  are the conventional Dirac matrices for the two fermions, which arise due to the scalar nature of the fermion-Higgs interaction (our notation for spinors presumes that $\langle u|\,\equiv \,u^\dag$).
Considering the retardation as a correction we expanded the amplitude in
powers of the transferred energy $\omega^2$. This energy and the transferred momentum 
${\bf q}$ are related to the fermion energies and momenta
$\omega\,=\,\varepsilon_1^\prime-\varepsilon_1\,=\, \varepsilon_2-\varepsilon_2^\prime$,
${\bf q}\, =\,{\bf p}_1^\prime -{\bf p}_1\,=\, {\bf p}_2-{\bf p}_2^\prime$.

As was mentioned, the term $\propto \omega^2$ in Eq.(\ref{Int1}) represents a correction that describes the retardation. Neglecting in this correction the relativistic effects we can use in this term the approximation $\beta_i\approx 1 $, $i=1,2$. We also can apply the nonrelativistic approximation relating $\omega$ 
to the fermion momenta
\begin{equation}
\omega\approx \frac{({\bf p}_1+{\bf q})^2-{\bf p}_1^2}{2m}=\frac{{\bf p}_2^2-
({\bf p}_2-{\bf q})^2}{2m}
\label{omegap1p2}
\end{equation}
Consequently we can rewrite (\ref{Int1}) as follows
\begin{equation}
\label{Int2}
V({\bf q})\approx
-g^2_h\left(\frac{\beta_1 \beta_2}{{{\bf q}^2 +m_h^2}} - \frac{({\bf q}^2 +2 {\bf p} \cdot 
{\bf q})^2}{4m^2({\bf q} ^2 +m_h^2)^2}\right)
\end{equation} 
Here we use the center of mass reference frame, ${\bf p}={\bf p}_1=-{\bf p}_2$, which is natural for the bound state problem addresses below.

The second term in the brackets in  Eq.(\ref{Int2}) is the contribution of the retardation to the relativistic correction. Let us simplify now the main, first term in the brackets in Eq.(\ref{Int2}) using the nonrelativistic approximation. With the help of the Foldy-Wouthuysen transformation we can relate the four-spinors $u$ with the nonrelativistic two-spinors $w$
\begin{equation}
u\,=\,\left( 
\begin{array}{c}
\big( 1-\frac{1}{8m^2}{\bf p}^2  \big) \,w \\
\frac{1}{2m}\,{\bm \sigma} \cdot {\bf p} \,w 
\end{array}
\right)
\label{uw}
\end{equation}
As a result we can use
instead of the matrices $\beta_1,\,\beta_2$  the following expressions
\begin{equation}
\beta_i\approx 1- \frac{ ({\bf p}+{\bf q}/2)^2}{2m^2}-
i\,\frac{ ({\bf q}\times {\bf p})\cdot {\bm \sigma}_i }{4m^2}~,
\label{beta-expansion}
\end{equation}
deriving after that
\begin{equation}
\beta_1 \beta_2\approx 1- \frac{ ({\bf p}+{\bf q}/2)^2}{m^2}-
i\,\frac{ ({\bf q}\times {\bf p})\cdot {\bm s}}{2m^2}~.
\label{beta-prod}
\end{equation}
Here 
\begin{equation}
{\bf s}=\frac12\,({\bm \sigma}_1+{\bm \sigma}_2)~.
\label{spin}
\end{equation}
is the operator of spin of the fermion pair.

Combining Eqs.(\ref{Int2}),(\ref{beta-expansion}) we find
\begin{align}
\nonumber
\!\!V({\bf q})\!=&-\frac{ g_h^2} { {\bf q}^2+m_h^2 }\Bigg(\!1+\frac{3m_h^2}{4m^2}
- \frac{ {\bf p}^2\!+2{\bf p}\!\cdot\!{\bf q} }{m^2}
-i\,\frac{ ({\bf q}\times {\bf p})\!\cdot \!{\bm s}}{2m^2}
\\
&-\frac{m_h^4-4 m_h^2 ({\bf q}\cdot {\bf p})+4({\bf q}\cdot{\bf p})^2}{4m^2({\bf q}^2+m^2)}\Bigg) +\frac{g_h^2}{2m^2}~.
\label{V(q)}
\end{align}
In order to apply this result to the bound state problem we identify here ${\bf p}$ with an operator of the momentum ${\bf p}= -i \bm{\nabla}$ and fulfill the Fourier transformation over the transferred momentum ${\bf q}$. The coordinate ${\bf r}$, which appears as the result of this transformation, should be kept on the left side of the operator ${\bf p}$.
This machinery is very similar to the one, which is conventionally applied for the derivation of the Breit interaction in QED, see Section 83 in \cite{QED} .
As a result after straightforward calculations we  find the following potential, which includes all relativistic and retardation corrections for two fermions interacting via the Higgs boson exchange 
\begin{align}
&W_\text{rel}\,=\,\delta T_\text{rel}+V_\text{rel}~,
\label{Wrel}
\\
&\delta T_\text{rel}\,=\,-\frac{p^4}{4m^3}~,
\label{delta T}
\\
&V_\text{rel}\, =\, \frac{2 \pi \alpha_h }{m^2} \,\delta(\bf{r})
\label{Vr}
\\
  - \frac{\alpha_h }{m^2 r}&
\exp{(-m_h \,r)}\left(\frac34 m_h^2\big(1-\frac{m_h r}{6}\big)
 + \Xi + \Lambda\right),
\nonumber	
\end{align}
where
\begin{align}
\Xi=&\big(1-\frac{m_h r}{2}\big)\frac{\partial^2}{\partial r^2} +\Big(1-2 m_h r + \frac{m_h^2 r^2}{2}\Big)\frac1r \frac{\partial}{\partial r},
\label{D}
\\
\Lambda=&\frac{(1+m_h r) \,{\bf l \cdot s} -3l(l+1)}{2r^2}.
\label{L}
\end{align}
It is presumed here that the state possesses the orbital momentum $\mathbf l$ and spin $\mathbf s$, and consequently the $\nabla$-operator was transformed into expressions with $\partial/\partial r$.
The first term $\delta T_\text{rel}$ in (\ref{Vr}) describes the conventional
relativistic correction to the kinetic energy, while $V_\text{rel}$
originates from $V(\bf q)$.
This term is written in a non-unitary form, which is similar to the way the Breit correction 
is often presented in QED, 
though in both cases this fact produces no complications in the first order of the perturbation theory. However, if necessity arises the potential can be amended to make the unitarity transparent, $V_\text{rel}\rightarrow V_\text{rel}^\prime=(V_\text{rel}+V_\text{rel}^\dag)/2$.
Note that the term $\Lambda$ vanishes for zero angular orbital momentum $l=0$. 
It is also interesting to note that the coefficient $\alpha_h/m^2=(4 \pi v^2)^{-1}$ in front of $\delta(\bf{r})$ and $\exp{(-m_h \,r)}$  terms in Eq. (\ref{Vr}) do not contain the mass $m$.

Remember that deriving $V({\bf q})$ in (\ref{V(q)}) we treated the retardation correction, which stems from the second term in the brackets in (\ref{Int1}),
by using the most simple nonrelativistic approach and also rewrote the first term 
from the brackets in (\ref{Int1}) using the Foldy-Wouthuysen transformation.
However for some applications it may be useful to treat the problem using the relativistic four-spinors.
The calculations presented in Appendix A show that with this purpose $V({\bf q})$ can be presented as follows
\begin{equation}
V({\bf q})=-{g_h^2}\Bigg[\frac{\beta_1 \beta_2}{{ {\bf q}^2+m_h^2}}+
\frac{ 
\big(\bm{\gamma}_1\!\cdot\!(2{\bf p}+ {\bf q})\big) 
~\big(\bm{\gamma}_2\!\cdot\!(2{\bf p}+ {\bf q} )\big) 
}
{ ({\bf q}^2+m_h^2)^2}
\Bigg]
\label{V-4spinors}
\end{equation}
We verified that when  the Foldy-Wouthuysen transformation is fulfilled to replace the Dirac
$\beta$ and $\bm{\gamma}$ matrices by their nonrelativistic approximations then Eq.(\ref{V-4spinors}) is reduced to Eq.(\ref{V(q)}). Note though that an attempt to fulfill the Fourier transform over ${\bf q}$ directly in Eq.(\ref{V-4spinors})  should be treated with care since the matrix elements of $\bm{\gamma}$ matrices in (\ref{V-4spinors}) effectively depend on ${\bf q}$. However, we will
not dwell on this issue here since Eqs.(\ref{V(q)}),(\ref{Vr}) and (\ref{V-4spinors}) constitute a reliable basis to treat the relativistic effects.

\subsection{Variational approach}
For the $S$ state, in which $l=0$, the last term in Eq.(\ref{Vr}) is absent, $\Lambda=0$. 
Taking the variational wave function $\psi$ from (\ref{psi}) 
we find the expectation value for the relativistic correction from  Eq.(\ref{Wrel})
\begin{equation}
\langle W_\text{rel}\rangle\equiv 
\langle\,\psi\,| W_\text{rel}|\,\psi\,\rangle=
\frac{q^4}{m^3}\left(-\frac54+\frac{6 \alpha_h m}{2q+m_h} \right).
\label{Vrq}
\end{equation}
Here the first and second terms in the brackets originate from the kinetic correction
$\delta T_\text{rel}$ and the potential correction $V_\text{rel}$ respectively.
The condition defining $q$ and the corresponding nonrelativistic approximation 
for the energy $E_0(q)$ were presented in Eqs.(\ref{Ea}), (\ref{qa}). The total energy $E$ can be found from $E=E_0(q)+\langle W_\text{rel}\rangle$.

It is instructive to present this correction in relation to the kinetic energy 
$T=q^2/m$
\begin{equation}
\frac{\langle W_\text{rel}\rangle}{T}=\frac{q^2}{m^2}\left(-\frac54+\frac{6 \alpha_h m}{2q+m_h} \right)
\label{VrT}
\end{equation} 

\subsection{Relativistic corrections to critical mass}

Taking the critical condition for the formation of the bound state 
from Eqs.(\ref{mh/2}), (\ref{mh}) we find that the right-hand side in (\ref{VrT})
reads
\begin{equation}
\frac{\langle W_\text{rel}\rangle}{T}=\frac{19}{64\pi^{2/3} }\left( \frac{m_h}
{ v}\right)^{4/3}\approx 0.042~.
\label{VrT2}
\end{equation} 
Here the value $m_h=100$ GeV was adopted in the last identity.

Obviously this small correction should result in a small variation of the critical mass.
Let us find it, but first note that the considered relativistic correction represents only one of the different perturbations, which influence the critical mass.
It is worth therefore to present an important simple formula for the
shift of the critical mass in general case, for an arbitrary perturbation
$\delta V$. Within the variational approach the necessary equation reads
\begin{equation}
\frac{\delta m_\text{cr}}{m_\text{cr}}\,=\,-\left(\frac{\partial E_0}{\partial m} \right)^{-1}\!
\frac{\langle \delta V\rangle}{m_\text{cr} }\,=\,
\frac{\langle \delta V\rangle}{3T}~.
\label{dm}
\end{equation}
Here $E_0(m)$ is the variational energy calculated without any corrections, while
$\langle \delta V\rangle$ is the perturbation potential averaged over the variational 
wave function.
To justify the first equality in Eq.(\ref{dm}) one expands the condition 
for the critical mass
$E_0(m_{cr}+\delta m_\text{cr})+\langle \delta V\rangle=0$ over $\delta m_\text{cr}$ in the vicinity of the mass $m_\text{cr}$, which satisfies the nonperturbed critical
condition $E_0(m_{cr})=0$.
To prove the last identity in Eq.(\ref{dm}) one 
differentiates over $m$ the energy $E_0$ specified in Eq.(\ref{Eh}).
Remembering that $\alpha_h=m^2/(4\pi v^2)$ is a function of $m$, one finds
\begin{equation}
\frac{\partial E_0}{\partial m}\,=\,
-\frac{q^2}{ m^2 } - \frac{8 \,\alpha_h \,q^3}{m\,(m_h+2q)^2}=-3\frac{q^2}{ m^2 }~,
\label{E0/m}
\end{equation}
where the last identity is derived from the critical condition $E_0=0$.
Thus one concludes that $\frac{\partial}{\partial m}E_0=-3T$, which justifies the last identity in (\ref{dm}).

Return now to the case of the relativistic correction.
Substituting the mentioned energy derivative into Eq.(\ref{dm}), specifying there that 
the perturbation is due to relativistic effects, $\delta V\rightarrow V_\text{rel}$, 
we find the relative increase of the critical mass due to the relativistic and retardation
corrections
\begin{equation}
\frac{\delta m_\text{cr}}{m_\text{cr}}\,=\,\frac{\langle \delta W_\text{rel}\rangle}{3T}\approx 0.014~.
\label{dmRad}
\end{equation}
In the last identity here Eq.(\ref{VrT2}) was used.
In absolute units $\delta m_\text{cr} \approx 0.014 m_\text{cr} \approx 7$ GeV.

We will see that the calculated in the next section radiative corrections  reduce this 
mass shift by 30\%,  resulting in the following total mass shift $\delta m_\text{cr} \approx 0.09 m\approx 5$ GeV. More accurate numerical results are mentioned below.

\subsection{Relativistic corrections for $m \gg m_h$}

    The relativistic correction rapidly increases with mass $m$. 
    For large masses  $m \gg m_h$ where $q \approx ({\tilde \alpha_h} +a) m/2$ approaches $q= \alpha_h m/2=m^3/(8 \pi v^2)$ we obtain
\begin{equation}
\frac{\langle W_\text{rel}\rangle}{T} \approx1.2 {\tilde \alpha_h}^2
\approx \Big(\frac{m}{3.4 v}\Big)^4 \approx \Big(\frac{m}{0.84 \, \mathrm{TeV}}\Big)^4
\label{VrT3}    
\end{equation}
Thus, the naive perturbative treatment of the relativistic corrections based on $W_\text{rel}$ 
from Eq. (\ref{Wrel}) definitely fails for $m>0.8$ TeV. 

Moreover, the more accurate numerical treatment shows that
this correction may be significant even at smaller masses.
This perturbation has a positive sign, 
as can be picked up from the positive
term $\propto \delta({\bf r})$ in Eq.(\ref{Vr}), which gives significant  contribution 
($\sim 60\%$). The repulsion produced by this correction may be so strong that it destabilises the bound state.

There is though the way around this difficulty. It will be discussed below, in Section 
\ref{Perturbation theory} in detail. Speaking briefly, the idea is to 
allow the wave function the freedom to adjust itself in such a way that
it is reduced at small separations $r$, where the relativistic correction is dominant.
This diminishing of $\psi(0)$ reduces the relativistic correction and makes the 
perturbative approach sound, and the bound state stable.

To prepare ourselves for this treatment of the relativistic correction we need to simplify Eq.(\ref{Vr}). With this purpose we, firstly, consider the most interesting for our 
discussion case of $S$-states, when the last term in this expression is absent, $\Lambda=0$. Secondly, take into account that the Higgs mass is relatively small, $m_h\ll m$. 
As a result, in the region of small separation between the 
two fermions $r\lesssim 1/m$ the Higgs mass is negligible.
Having this in mind we simplify the relativistic correction setting 
$m_h=0$ in the term $\Xi$, which gives
(\ref{Vr}) for 
\begin{equation}
\Xi\,=\,\frac{\partial^2}{\partial r^2} +\frac1r \frac{\partial}{\partial r}=
\Delta-\frac1r \frac{\partial}{\partial r}~.
\label{Dmh=0}
\end{equation}
Here the last identity takes into account that we consider $S$-states.
Hence we find that for $m_h=l=0$ Eq.(\ref{Vr}) gives
\begin{equation}
V_\text{rel}\, =\, \frac{r_0}{m} \,\Big(
2 \pi \,\delta({\bf r})
  +\frac1{r^2} \,\frac{\partial}{\partial r}-  \frac1r\, \Delta\Big)~.
\label{Vrelddd}
\end{equation}
To simplify notation the new parameter
\begin{equation}
r_0\,=\,\frac{\alpha_h}m
\label{r0}
\end{equation}
is introduced here (compare the classical radius of a fermion in QED, $r_\text{clas}=\alpha_\text{qed}/m$).
To simplify (\ref{Vrelddd}) let us treat 
the sum of the first and second terms there as a perturbation.
An inspiration comes from an identity
\begin{equation}
\langle \psi| \delta({\bf r})+\frac1{2\pi r^2} \frac{\partial}{\partial r}|\psi\rangle=
\psi^2(0)+
2\!\int_0^\infty \!\!\!\psi(r)\frac{d \psi(r)}{dr}dr=0,
\label{delta-psi}
\end{equation}
which implies that if one neglects these two terms from (\ref{Vrelddd}) 
and then contemplates the perturbation theory over 
them, one can be certain that
the first order of this perturbation is absent.
Neglecting the higher-order contribution of these terms 
(see discussion at the end of this subsection) 
we suppress these terms deriving from Eq.(\ref{Vrelddd}) 
\begin{equation}
V_\text{rel}\, \approx \, -\frac{r_0}{m} \,\frac1r\,\Delta~.
\label{Vreld}
\end{equation}
We can combine now this relativistic correction with the nonrelativistic Hamiltonian $H_0$ from (\ref{Hh}) and write the corresponding eigenvalue problem $E\psi=(H_0+V_\text{rel})\psi$. Multiplying this differential equation by the factor $r/(r+r_0)$  we present the result as follows
\begin{equation}
\frac r{r+r_0}\,E\,\psi\,=\,
\Big(\frac{{\bf p}^2}{m}-\frac{\alpha_h}{r+r_0}\,e^{-m_h\,r}\Big)\,\psi
\label{SL}
\end{equation}

Clearly Eq.(\ref{SL}) can be considered as the eigenvalue problem of the Sturm-Liouville  type. It has a very clear structure, but can be simplified further. Take into consideration
that for not too large $m$, say $m\lesssim 1$ TeV, $r_0$ is expected 
to be smaller than the typical size of the bound state.
Hence we can use the simplification $r/(r+r_0)\approx 1$ for the factor on the left-hand side of (\ref{SL}). For large separation $r\gg r_0$ the validity of this approximation is obvious. In the opposite case of small distances $r\lesssim r_0$ the term with $E$ is much smaller
then the potential energy and can be ignored in the differential equation anyway. Thus the replacement $r/(r+r_0)\approx 1$ does not stir things up. 

We conclude that the wave function, which describes the binding of two heavy fermions satisfies the 
Schr\"{o}dinger-type eigenvalue problem $E\,\psi\,=\,H_{0+\text{rel}}\,\psi$, 
in which the Hamiltonian
\begin{equation}
H_{0+\text{rel}}\,=\,\frac{{\bf p}^2}{m}-\frac{\alpha_h}{r+r_0}\,e^{-m_h\,r}
\label{rs}
\end{equation}
accounts for the conventional nonrelativistic interaction and 
takes into account the relativistic and retardation corrections. 

A notable simplicity and clear physical nature of this result make it attractive. The only distinction of (\ref{rs}) from the nonrelativistic Hamiltonian (\ref{Hh}) lies in the factor $r_0$, which effectively cuts the Yukawa-type potential off at small distances. As a result at small distances there arises an effective repulsion, which is in line with the repulsive nature of the relativistic correction mentioned previously.

It is worth to summarize the set of approximations employed for the derivation of (\ref{rs}). First, the Higgs mass was neglected in the relativistic term. Second, the approximation $r/(r+r_0)\approx 1$ on the left-hand side of (\ref{SL}) was used. Both these 
approximations are justified by the parameters of the problem. 
Third, the sum of the first two terms  in Eq.(\ref{Vrelddd}) was neglected. We can 
see now the qualitative reason, which  makes this approximation sound. The perturbation theory over these two terms starts from only the second order, as (\ref{delta-psi}) shows. Meanwhile the repulsion at small separations caused by the relativistic correction is well accounted for in the Hamiltonian $H_{0+\text{rel}}$. This repulsion reduces the wave function at small distances and hence suppresses the impact of the omitted terms in the problem. Consequently, their contribution is to be suppressed. This implies, in particular that the sign of the correction to (\ref{rs}) is defined by the second order perturbation theory 
(over the omitted in (\ref{Vrelddd}) terms), and hence is definitely negative.

 \section{radiative corrections}
   \label{rad}
\begin{figure}[tbh]
  \centering \includegraphics[ height=3cm, keepaspectratio=true, angle=0]{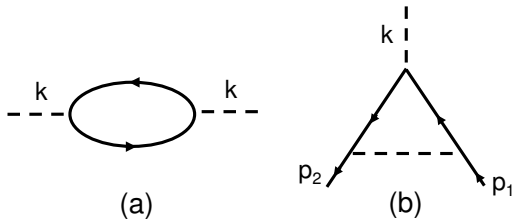}
\vspace{0cm}
\caption{Two most important radiative processes: (a) - the vacuum polarization
produced by heavy fermions, which affects propagation of the Higgs boson,  and 
(b) - the vertex correction for the Higgs-fermion interaction;
solid and dashed lines - fermion and Higgs fields.
}\label{fig2} 
   \end{figure}

 To calculate the radiative corrections we use the conventional method which originally was developed to calculate the Lamb shift in atoms - see e.g. \cite{QED}. 
For the problem at hand the necessary corrections were 
calculated in \cite{Kuchiev-1012.0903}. 
It was shown there that the most important contribution give
two diagrams, shown in Fig. \ref{fig2}. 
One
describes the Higgs polarization operator produced by the fermion loop,
another the vertex correction to the Higgs-fermion interaction.

   \noindent

Note that the diagram with the virtual fermion annihilation does not contribute to the energy shift for the non-relativistic fermions since the annihilation vertex of the fermion and anti-fermion into the Higgs  vanishes for $p^2=0$. The reason is that the Higgs is a scalar. This is in contrast to a similar annihilation processes with the virtual annihilation of the positronium into photon, vector particle, which gives a nonzero contribution in QED \cite{QED}. We also neglect the diagram with crossed Higgs legs in the two-Higgs exchange since it is not enhanced by a large number of different fermions ($\nu$)  in the loop (compare with the polarization operator  Eq. (\ref{P})).  

\subsection{Polarization operator}
In the approximation $k\ll m$ and $m_h \ll m$ the polarization operator represented by the diagram (a) in Fig. \ref{fig2}) equals \cite{Kuchiev-1012.0903}  
\begin{equation}
P(k^2)\approx\,-  \nu\,\frac{\,g_{h}^2\,(k^2-m_h^2)^2 }{ 80\pi^2m^2 }=
-\frac{\nu}{8\pi^2}\,\frac{(k^2-m_h^2)^2}{10 v^2}.
\label{P}
\end{equation}
Here $\nu$ is the number of heavy fermions with mass $m  \gg m_h$ (recall the factor $N_c=3$ for quarks). Importantly, the fermion mass $m$  is canceled out in the final expression here since $g_{h}=m/v$.
Hence the contributions of all heavy fermions are  summed up. For one heavy lepton $\nu=1$, for 
the whole 4-th generation $\nu=8$, and $\nu=11$ when the top quark is also counted. We neglect the contributions of W$^\pm$,  Z$^0$ and Higgs in the loop since condition $m \gg m_h$ is not valid for them. 
The radiative correction to  the Higgs potential $V_h$ is given by 
\begin{equation}
\delta V^\text{P}_h(k)=\frac{P(k^2)V_h(k)}{k^2-m_h^2}\approx-  \nu \frac{g_h^2(k^2-m_h^2) V_h(k)}{ 80\pi^2m^2 }
\label{PVh}
\end{equation}
In the coordinate representation this gives a potential proportional to the $\delta$-function
(similar to the radiative correction to the energy in the positronium or hydrogen).

\subsection{Vertex}

Consider now diagram (b) in Fig.\ref{fig2}, which describes correction $\Gamma$ to the vertex 
 of the fermion-Higgs interaction. The expansion of $\Gamma(k)$ in powers of $k^2/m^2$ gives \cite{Kuchiev-1012.0903} 
\begin{equation}
\frac{1}{g}~\Gamma(k^2)\,\approx \,1+\frac{g_h^2}{8\pi^2}\,\gamma\,\frac{k^2}{m^2}
\,=\,1+\frac{\gamma}{8\pi^2} \,\frac{k^2}{v^2}~,
\label{fT}
\end{equation}
where $\gamma$ is an expansion coefficient, which was calculated in \cite{Kuchiev-1012.0903}. 
The found there  analytical expression for $\gamma$ is lengthy, but a simple fitting proposed in this work,
\begin{equation}
\gamma\approx \gamma_\text{\tiny{fit}}=\frac{1}{3}\,\Big(\ln \frac{m+m_h}{m_h}-
\frac{7m}{4m+5m_h} \Big)~,
\label{fit}
\end{equation}
reproduces $\gamma$ with per cent accuracy. 
The correction to the Higgs-fermion vertex constant is 
$\delta g_h= \Gamma(k^2)-g_h$. We find from this the correction to the Higgs exchange potential
\begin{equation}
\delta V^\Gamma_h(k)=\frac{ 2\,\delta g_h}{g_h}V_h(k)=\frac{\alpha_h \gamma \,k^2 V_h(k)}{\pi \,m^2}
\label{GVh}
\end{equation}

\subsection{Radiative potential and correction to energy}
Summing up the contributions of the polarization operator $\delta V^\text{P}_h$ and the vertex $\delta V^\Gamma_h$ 
we find for them the following result, which is convenient to deal with using the coordinate representation
\begin{equation}
V_\text{rad} ({\bf r})=\frac{4 \alpha_h}{m^2}\Big( 
\alpha_h\big(\gamma - \frac{\nu}{20}\big) \,\delta({\bf r}) +\gamma \,m_h^2 \,V_h(r)\Big)
\label{Vrad}
\end{equation}
Using the variational wave function $\psi(r)$ from 
Eq.(\ref{psi}) we obtain the radiative correction to the ground state energy
\begin{equation}
\langle V_\text{rad} \rangle =\frac{4 \alpha_h^2 q^3}{\pi m^2}\Big(\gamma - \frac{\nu}{20} - \frac{4 \pi \gamma  m_h^2}{(m_h +2q)^2}\Big)
\label{Erad}
\end{equation}
For $q=q_\text{cr}=m_h/2$ and the number of heavy fermions in the polarization loop  $\nu=11$ we obtain the ratio of the radiative correction to the kinetic energy 
\begin{equation}
\frac{\langle V_\text{rad} \rangle}{T} \approx -0.015
\label{EradT}
\end{equation}
We see that the radiative correction  is about 3 times smaller than the relativistic correction 
Eq. (\ref{VrT2})  and is opposite in sign. For higher mass $m$ there are  significant cancellations between  different terms in the radiative correction Eq. (\ref{Erad}).  As a result  the radiative correction at $m\sim 800$ GeV is about 5 times smaller than the relativistic correction and has the opposite sign.

\section{Variational energy at large masses}

\subsection{Relativistic kinetic energy}
We used previously the perturbation theory 
to account for the relativistic correction $-p^4/(4m^3)$ to the kinetic energy.
This can be improved if we calculate explicitly the matrix element
of the relativistic kinetic energy 
$T_\text{rel}=2\langle (m^2 +p^2)^{1/2}-m\rangle$
on the probing wave function (\ref{psi}), which can be written as follows
\begin{equation}
\langle T_\text{rel}\rangle =2\int\Big((m^2 +p^2)^{1/2}-m\Big)\,\psi^2(p) \,\frac{d^3 p}{(2\pi)^3}~.
	\label{Trel0}
\end{equation}
where $ \psi (p)=8\, \pi^{1/2} \,q^{5/2}\,(q^2+p^2)^{-2}$ is the Fourier transform of the wave function (\ref{psi}).  Straightforward  integration yields
\begin{equation}
\frac{\langle T_\text{rel}\rangle}{m}=
\frac{4}{\pi}\Big(
\frac{ x\,(3-4x^2+4x^4) }{3(1-x^2)^2} +
\frac{(1-2x^2) \arccos x }{ (1-x^2)^{5/2} }\Big)-2\quad
	\label{Trel}
\end{equation}
where $x=q/m$. 
The asymptotes of this expression are $\langle T_\text{rel}\rangle/m\approx x^2(1-5x^2/4+128x^3/15)$ when $x\rightarrow 0$
and $\langle T_\text{rel}\rangle/m\approx 16\,x/(3\pi)-2$ for $x\rightarrow \infty$.
One verifies also that a very simple interpolating formula $\langle T_\text{rel} \rangle\approx m x^2/(1+0.3 x^2)$ reproduces the accurate expression for $\langle T_\text{rel} \rangle$ with errors below 12\% provided $q\le 2m$.
Equation (\ref{Trel}) for the kinetic energy allows one to include the relativistic corrections into the variational procedure outlined below. This contrasts
the naive perturbative correction $-p^4/4 m^3$, which negative sign results in an  unlimited increase of the momentum $p$ during the minimization of the energy. 

\subsection{Improved variational energy}

We can  formulate now an improved version of the variational energy,
which incorporates the relativistic, retardation, long-range, and radiative corrections
\begin{equation}\label{Eimproved}
E(q)= \langle T_\text{rel}\rangle  + \langle V_\text{Y}\rangle - a q
+\langle V_\text{rel}\rangle
+\langle V_\text{rad}\rangle\,.
\end{equation} 
Here all terms on the right-hand side are functions of $q$.
The first term $\langle T_\text{rel}\rangle $ defined in (\ref{Trel}) represents the kinetic energy modified by the kinematic relativistic correction.
The second one takes into account the Yukawa-type attraction, 
\begin{equation}
\langle V_\text{Y}\rangle\,=\,-\langle \psi |\,(\alpha_h/r)\, e^{-m_hr}\,|\psi\rangle=- \frac{4 \alpha_h q^3}{(m_h+2q)^2}~,
\label{yuk}
\end{equation}
being identical to the second term from Eq.(\ref{Eh}).
The term $-aq$ comes from the long-range potential, gluon or photon exchange.
The term $\langle V_\text{rel}\rangle$ 
incorporates the relativistic corrections and the effect of retardation.
Note that currently we include the kinematic relativistic correction
into $\langle T_\text{rel}\rangle $. In our previous arrangements 
the kinematic relativistic correction was taken into account in $\langle \delta T_\text{rel}\rangle$
via Eq.(\ref{delta T}). This correction shows itself as the first term in the brackets in Eq.(\ref{Vrq}), $\langle T_\text{rel}\rangle=-5q^4/(4m^3)$. 
We do not need to include this term here since it is 
covered by $\langle T_\text{rel}\rangle $.
The second term in the same brackets in (\ref{Vrq}) represents the correction to the potential, which stems from the relativistic and retardation effects and contributes to Eq.(\ref{Eimproved})
\begin{equation}
\langle V_\text{rel}\rangle\,=
\,\frac{6\, \alpha_h q^4}{m^2(2q+m_h)}~.
\label{Vdyn}
\end{equation}
The last term $\langle V_\text{rad}\rangle$ in (\ref{Eimproved}) represents the radiative correction calculated in Eq. (\ref{Erad}). 
It is suppressed by a small factor $\alpha_h/\pi$ in comparison to the relativistic correction. 
All five terms on the right-hand side of Eq.(\ref{Eimproved}) are functions of the
variational parameter $q$. Minimizing the total energy $E(q)$ over $q$, one can find the variational estimate for the ground state of the two fermions. Numerical results are discussed below.

\section{Numerical data}

\label{Calculations}

\subsection{Perturbation theory}
\label{Perturbation theory}

To evaluate and study the behavior of the ground state energy 
for two heavy fermions bound together we employ
the conventional approach based on
the Schr\"{o}dinger - type  eigenvalue problem
\begin{equation}
E \psi(r)\,=\,H\psi~.
\label{H1+deltaV}
\end{equation}
We divide the total Hamlitonian
into two parts, $H=H_1+\delta V$,
and solve the  Schr\"{o}dinger equation (\ref{H1+deltaV}) accurately
for $H\approx H_1$ while applying the first order of the 
perturbation theory to the term $\delta V$.

In problems of this type the term $\delta V$ often
accounts for the relativistic and radiative corrections,
which is the case in many atomic and nuclear problems.
However, in the problem at hand we are facing 
a surprising deviation from this rule, 
the relativistic correction turns very large and repulsive.
The repulsion is so strong that it is difficult to formulate the problem
by choosing the naive Yukawa-type approximation, which ignores this
repulsion at the initial step. 
If one tries to implement this naive approach and identifies
$H_1\equiv H_0$, where $H_0$ (\ref{Hh}) includes the pure Yukawa potential $V_\text{Y}(r)$ from (\ref{Vh}), then the bound state found on the first step of such analyses
would be destabilized, pushed into the continuum by the first order correction over the relativistic potential $\delta V\equiv W_\text{rel}$ (\ref{Wrel}).

To overcome this difficulty let us have in mind that
the relativistic correction is large predominantly in the 
region of small separation  $r$ between the fermions. 
The difficulty introduced by the naive Yukawa-type
interaction is that it makes the wave function too
large at the origin. As a result, the repulsion produced by
the relativistic corrections gains strength.

The necessary remedy is clear. One needs to choose the initial approximation for $H_1$, 
which allows the wave function to be further extended in the region of large $r$, and correspondingly reduced at small $r$. Having this in mind we include into 
the Hamiltonian $H_1$ the modified Yukawa interaction from (\ref{rs})
\begin{equation}
H_1\,=\,H_{0+\text{rel}}\,=\,\frac{{\bf p}^2}{m}-\frac{\alpha_h}{r+r_0} e^ {-m_h\, r}~.
\label{H11}
\end{equation}
Compared with the pure Yukawa interaction, its modified version 
used here is not singular at the origin, and hence 
produces an effective repulsion 
at small $r$. 
Consequently it pushes the wave function 
out of the origin.
An additional advantage of this approximation is that
it has a clear physical meaning accounting
for the relativistic correction, albeit in the simplified form.

Eq.(\ref{H11}) neglects the long-range potential $V_\text{lr}(r)$ in (\ref{Vl}), which
stems from the possible gluon and photon exchange. The photon induced interaction
should be considered as a small perturbation. The neglect of the gluon exchange is
obviously justified for leptons. For the quarks the problem is addressed
in subsection \ref{Gluon-exchange}.

In the first order of the perturbation theory
we subtract the approximate form for the interaction introduced in (\ref{H11}), adding instead the necessary 
pure Yukawa-type potential and also adding 
the accurate relativistic potential
\begin{align}
&\delta V = V_\text{rel, res}+V_\text{rad}~.
\label{WVrad}
\\
&V_\text{rel, res}=W_\text{rel}-\alpha_h\left(\frac{1}{r}-\frac{1}{r+r_0}\right) \, e^ {-m_h\, r}
\label{residual}
\end{align}
Here the term $V_\text{rel, res}$ can be called the residual relativistic 
correction. It includes the proper relativistic correction
$W_\text{rel}$ from (\ref{Wrel}) and the additional term 
$\propto \alpha \exp{(-m_h\, r)}$, which
include the difference between the
modified Yukawa interaction and the proper Yukawa potential. 
On top of it, the radiative 
correction $V_\text{rad}$ from Eq.(\ref{Vrad}) 
is added into $\delta V$.

Let us show now that Eqs.(\ref{H1+deltaV})-(\ref{residual}) 
provide a sensible approach to the problem.
To do this we calculated the binding energy $\varepsilon$ of the 
pair of heavy fermions.
Fig. \ref{fig3} presents this energy versus the mass of the fermion. The negative energy
means that the bound state exists.
The Higgs mass in this calculation was chosen  $m_h=100$ GeV, and the number of fermions contributing to the loop in the polarization operator was taken $\nu =11$.

Observe first of all that the way the perturbation theory is formulated 
Eqs.(\ref{H1+deltaV})-(\ref{residual}) makes sense.
Compare the dotted and thick dashed lines in Fig.\ref{fig3}. The first one represents
the eigenvalue $\varepsilon_1$ derived from Eq.(\ref{H1+deltaV}) when $H\rightarrow H_1$.
The second shows  $\varepsilon = \varepsilon_1+\delta\varepsilon$,
where $\delta \varepsilon$ originates from the residual relativistic correction 
$V_\text{rel,res}$ from (\ref{residual}).
Note a reasonable agreement between the two sets of data, 
their discrepancy is $29-38\%$ for all the region shown,
which makes the perturbation theory applicable.
Hence, the accuracy of the found binding energy can be roughly estimated
as $\sim 15\%$.
The found residual relativistic correction is negative,
which is due to the negative sign of the second, additional term
$\propto \alpha \exp{(-m_h\, r)}$ in Eq.(\ref{residual}).
This contrasts the positive sign produced by 
the proper relativistic correction
$V_\text{rel}$ in this equation.

Importantly, the bound state is definitely
stable, it is not destabilized by the relativistic corrections. 
The calculations in Fig.\ref{fig3} demonstrate this
fact in the first order of the perturbation theory.
Moreover, if the second order of the perturbation theory over $V_\text{rel,res}$ 
would be implemented (we do not fulfill these calculations, only contemplate the expected outcome), then the negative contribution of the second-order correction 
would make the bound state only tighter.
Thus, the way the perturbation theory is introduced makes the relativistic correction manageable. This is in contrast to the naive formulation of the problem, 
in which the pure Yukawa-type potential is taken as a starting point, while the 
relativistic correction becomes a problem.

The total ground state energy, which is shown in Fig. \ref{fig3} by 
the thick solid line, includes also the radiative correction 
{calculated from $V_\text{rad}$ in Eq.(\ref{WVrad}). Observe that the latter is 
small and negative, compare the difference between the solid and dashed thick lines.

\begin{figure}[t]
  \centering \includegraphics[ height=5.2cm, keepaspectratio=true, angle=0]{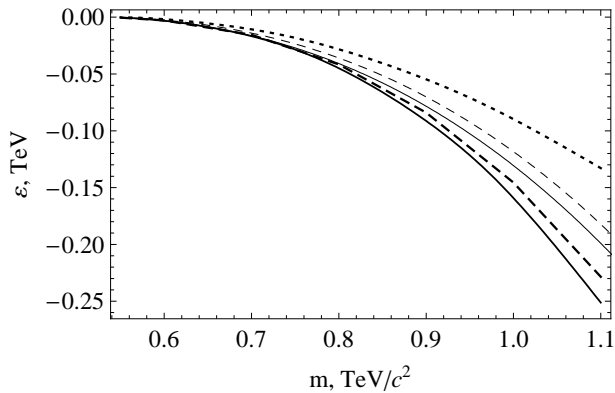}
\vspace{0cm}
\caption{
Binding energy $\varepsilon$ of two quarks versus the fermion mass $m$;
thick doted, dashed and solid lines are:
the eigenvalue of the Hamiltonian $H_1$ (\ref{H11}), the residual (see text) relativistic correction $\delta \epsilon=\langle V_\text{rel, res}\rangle$ (\ref{residual}) is added, the full correction $\delta \epsilon=\langle \delta V \rangle $, which includes the radiative correction is accounted for; thin lines: the variational energy $\varepsilon_\text{var}$ from (\ref{minEq}) with (solid) and without (dashed) the radiative correction;
$m_h=100$ GeV, $\nu=11$.
}\label{fig3} 
   \end{figure}
   \noindent

Let us find the critical mass of fermions, which guarantees 
an existence of the bound state. 
Fig. \ref{fig4}, which shows the rescaled version of our data,
allows us to state that $m_\text{cr}\approx 515$ GeV.
Remember that Eq.(\ref{mcr}) predicts $m_\text{cr}\approx 503$.
It was mentioned that the pure Yukawa-based approach is 
not reliable, that the relativistic repulsion is strong. Nevertheless, 
we have to resign to the conclusion that the naive estimate of the 
critical mass based on the Yukawa-type approximation Eq.(\ref{mcr}) 
is reasonably accurate.

\begin{figure}[tbh]
  \centering \includegraphics[ height=5.2cm, keepaspectratio=true, angle=0]{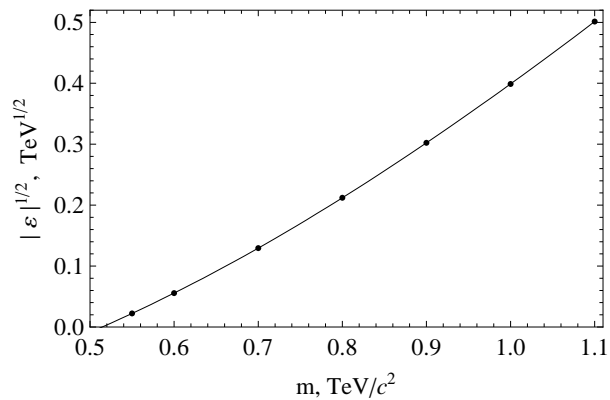}
\vspace{0cm}
\caption{
The square-root of the binding energy $(-\varepsilon)^{1/2}$ for two quarks 
versus the quark mass $m$, $\varepsilon$  is extracted from Eqs.(\ref{H1+deltaV})-(\ref{residual})
and includes the relativistic and radiative corrections (same data is 
shown by thick solid line in Fig. \ref{fig3});
dots - calculations, solid line - interpolation.
}\label{fig4} 
   \end{figure}
   \noindent

The discussed physical arguments and numerical results are self-consistent,
and hence look reliable. However, we are to keep in mind that the problem 
considered is unusual, the relativistic corrections 
are strong and repulsive.
It is fortunate therefore that there is the way to verify our results by 
independent calculations discussed in the next subsection.

\subsection{Variational approach}

The wave function of the nonrelativistic problem
has a very simple shape, possesses one maximum and shows no nodes. 
One should expect therefore that the simplest, one-parameter approximation for this function 
in Eq.(\ref{psi}) provides a steady basis for the variational approximation. 
Following this path we considered the energy $E(q)$ found in  Eq.(\ref{Eimproved})
as a function of the parameter $q$ introduced in the wave function (\ref{psi}). 
Minimizing this function one finds the binding energy 
\begin{equation}
\varepsilon_\text{var}\,=\,{\min}_q\,E(q)~.
\label{minEq}
\end{equation}
In accord with the previous discussion let us neglect the 
term in $-a q$ in $E(q)$, which accounts for the 
gluon exchange (it is discussed in some detail below). 
The found variational binding energy is shown in Fig. \ref{fig3} by two thin lines. 
The thin solid line takes into account the total energy from 
(\ref{Eimproved}), while 
thin dashed line neglects the radiative correction, i.e. discards the last term
$\langle V_\text{rad} \rangle $ in this equation.

Fig. \ref{fig3} shows that the two approaches to the problem,
one based on the perturbation theory, another on the variational approach,
are close. The discrepancy is below $21\%$ in all the region shown.
This discrepancy does not contradict the accuracy 
of the quantum mechanical calculation $\sim 15\%$.
It also  complies with the accuracy of the variational approach,
which in the problems of this type should better than $\sim 20\%$.
It is interesting that the two approaches also agree on
the 
corrections. Compare the radiative correction,
which is shown by the discrepancy of the solid and dashed lines,
(the two lines are thick for quantum mechanical and thin for
variational calculation). 
The observed agreement of two so differently formulated approaches is satisfying. 

\begin{figure}[tbh]
  \centering \includegraphics[ height=5.2cm, keepaspectratio=true, angle=0]{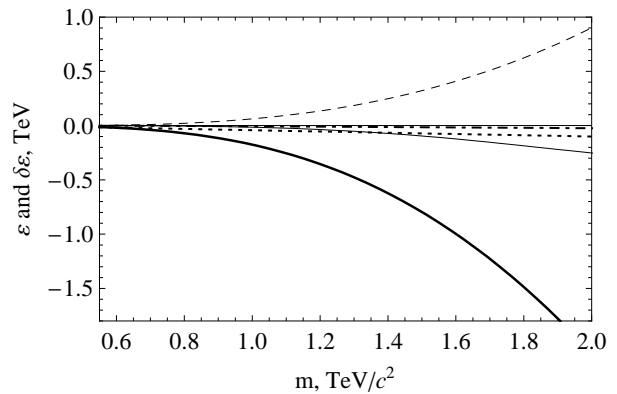}
\vspace{0cm}
\caption{From bottom to top: thick solid, thin solid, dotted, dash-dotted and dashed lines: 
total binding energy $\varepsilon$,
the radiative correction $\langle V_\text{rad}\rangle$,
the gluon correction $-a q$,
pure kinematic relativistic correction
$-p^4/(4m^3)$, 
and relativistic correction $\langle V_\text{rel}\rangle$  respectively, all calculated using the variational approximation based on Eq.(\ref{Eimproved});
$m_h=100$ GeV, $\nu=11$ (compare Fig.\ref{fig6}, which illustrates the case of $m_h=600$ GeV).
}\label{fig5} 
   \end{figure}
   \noindent

\begin{figure}[tbh]
  \centering \includegraphics[ height=5.2cm, keepaspectratio=true, angle=0]{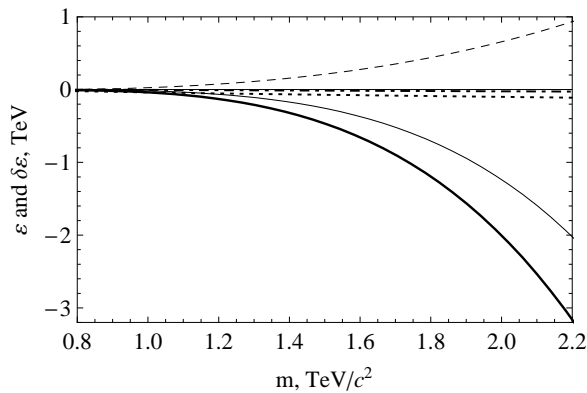}
\vspace{0cm}
\caption{From bottom to top: thick solid, thin solid, dotted, dash-dotted and dashed lines: 
total binding energy $\varepsilon$,
the radiative correction $\langle V_\text{rad}\rangle$,
the gluon correction $-a q$,
pure kinematic relativistic correction
$-p^4/(4m^3)$, 
and relativistic correction $\langle V_\text{rel}\rangle$  respectively, all calculated using the variational approximation based on Eq.(\ref{Eimproved});
$m_h=600$ GeV, $\nu=11$ (compare Fig.\ref{fig5}, which illustrates the case of $m_h=100$ GeV).
}\label{fig6} 
   \end{figure}
   \noindent

Fig. \ref{fig5} and Fig. \ref{fig6}  illustrate the role played by different perturbations in the problem
(Fig. \ref{fig5} is for $m_h=100$ GeV and  Fig. \ref{fig6} is for $m_h=600$ GeV) .
We start our discussion from  Fig. \ref{fig5}. 
The solid line there shows the total binding energy, which is same as 
in Fig. \ref{fig3}, and is repeated here to provide 
the general scale of energies.
The dashed and dotted lines show the relativistic
and radiative corrections, which are given by the terms 
$\langle V_\text{rel}\rangle$ and 
$\langle V_\text{rad}\rangle$ from Eq.(\ref{Eimproved}).
Note that the relativistic correction is positive and very large.
Remember that this point was an obstacle, which was overcome
using the approach to the perturbation theory formulated in the previous subsection.
(Note also that here the proper relativistic correction
$\langle V_\text{rel}\rangle$ is considered. 
It should not be confused with the residual interaction $V_\text{rel,res}$, which
arises in the perturbation theory, see
Eq.(\ref{residual}). As was mentioned, the latter is smaller and has the negative sign, see 
Fig. \ref{fig3}.)
It is interesting that the major contribution to the relativistic correction
gives the potential energy $\langle V_\text{rel}\rangle$ in Eq.(\ref{Eimproved}).The simple kinematic correction,
$-p^4/(4m^3)$, which is accounted for by the term $\langle T_\text{rel}\rangle$ in this equation proves to be very small, see the dot-dashed line.

The radiative correction shown in Fig. \ref{fig5} proves to be negative and small, note the blowing up factor ten used to make it visible. 
To emphasize the point we extended data up to very large fermion masses $m=2$ TeV and 
found that the correction remains very small even in this mass interval.
The found minor role of the radiative corrections in the two-fermion system 
complies with our previous results for the fermion bags \cite{Kuchiev-1012.0903,Kuchiev:2011}.
The situation for a large Higgs mass is different. The radiative corrections are significantly more
important there - see Fig. \ref{fig6}.

\subsection{Gluon exchange}
\label{Gluon-exchange}

The numerical examples presented previously neglected 
the processes with the gluon exchange, 
which is acceptable provided the two heavy leptons are considered.
However, for two heavy quarks the role of the gluon exchange needs to be examined.
For simplicity let us approximate the
impact of the gluon exchange by the long-range potential
$V_\text{lr}(r)$ from Eq.(\ref{Vl}). Then the relative strength
and impact of the gluon exchange
can be estimated by comparing the effective coupling constant
$a=4 \alpha_s/3$ in this potential with the Higgs coupling
$\alpha_h=m^2/(4\pi v^2)$. Taking for estimation
$\alpha_s=\alpha_s(M_Z)\approx 0.12$ one finds that 
the mass after which the Higgs exchange dominates is
$m_\text{h-dom}\approx 349$ GeV. Meanwhile, as was mentioned, 
the critical mass that makes the bound state due to the Higgs exchange possible 
is $m_\text{cr}\approx 515$ GeV.
Hence $\alpha_h$ at this mass substantially exceeds $\alpha_s$,
$\alpha_h(m_\text{cr})\approx \,2.2\, (4\,\alpha_s/3)$, 
becoming dominant at larger masses. 
For example, at $m=1100$ GeV they differ by an order of magnitude,
$\alpha_h(1100 \mathrm{\,GeV})\approx 10 \,(4\,\alpha_s/3)$.

\begin{figure}[tbh]
  \centering \includegraphics[ height=5.2cm, keepaspectratio=true, angle=0]{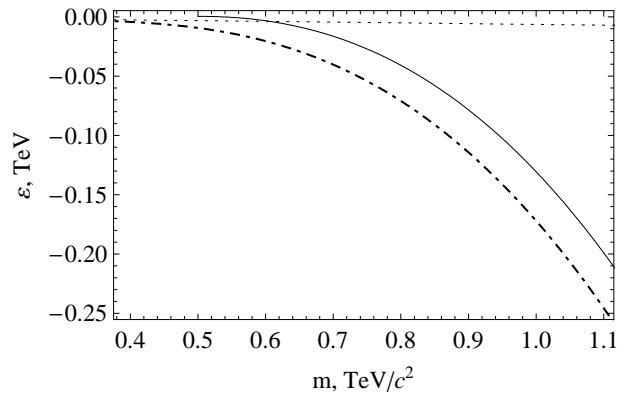}
\vspace{0cm}
\caption{
The gluon exchange in the binding energy of two quarks; 
solid - Higgs exchange only, the energy
$\varepsilon$ with the relativistic and radiative corrections (same as the thin solid line in Fig. \ref{fig3}),
dot-dashed - additionally,
the gluon contribution into the energy is included,
dotted - naive Coulomb-type estimate of the binding energy produced by gluons; 
calculations are based on the variational approach; $m_h=100$ GeV, $\nu=11$.
}\label{fig7} 
   \end{figure}
   \noindent

One concludes from the mentioned numbers that the gluon
exchange plays a role of a perturbation, but the one which may give 
a sizable contribution. In order to calculate it we use the variational approach
based on Eq.(\ref{Eimproved}), in which we include now the 
contribution of the long-range potential $\langle V_\text{lr}\rangle=-\frac43 \,\alpha_s q$
(omitted in our previous discussion).

The results found for the binding energy are shown in Fig. \ref{fig7}.
Several engaging features can be observed there.
Firstly, the discrepancy between the binding energy calculated with
and without the gluon exchange is quite sizable.
Secondly, the relative discrepancy between these binding energies is obviously
falling with the fermion mass, 
which agrees with the fact that the ratio $\alpha_s/\alpha_h$ is diminishing.
However, their absolute discrepancy remains approximately constant.
Thirdly, it is interesting that the shift of the binding energy
produced by the gluon exchange greatly exceeds the typical binding energy
observed when the Higgs exchange is neglected.
In the latter case one can roughly estimate the spread of binding 
energies due to the gluon 
exchange alone in the Coulomb-type model $\varepsilon_g\approx\varepsilon_c= -\frac{m}{4}\,(\frac43\,\alpha_s)^2$. 
Fig.  \ref{fig7} shows that this energy is much smaller than the shift of the binding energy when the gluon exchange is added to the Higgs exchange. 

There is a  clear  physical reason behind this phenomenon. The gluon exchange alone
keeps two quarks at a typical separation that is comparable with the effective Coulomb
radius. In contrast, the Higgs exchange is able to bring the quarks much closer,
which makes the gluon exchange more effective.

A similar physical phenomenon takes place in the region $m\le m_\text{cr}$.
Here the Higgs exchange alone is not able to produce the bound state, but when
combined with the gluon exchange it increases the binding energy significantly.

\section{Analytical solution at small Higgs mass}

For a wide range of fermion masses discussed in the data
presented in Figs. \ref{fig3}-\ref{fig7} both the fermion mass and typical momenta
of quarks in the bound state exceed the Higgs mass. 
Consequently it makes sense to investigate the limit $m_h=0$. 
Moreover, the same data shows that the 
radiative correction and the relativistic kinematic correction
are small, see Fig. \ref{fig7}.
In the simplest approximation we can therefore safely neglect these corrections.
Making these simplifications in Eq.(\ref{Eimproved}) we find a clear analytical expression for the energy
\begin{equation}
\label{Eanalyt}
E(q)\approx E_a(q) =
\frac{ {q}^2}{m}  - \alpha_h{q}+ 3\,\frac{\alpha_h{q}^3 }{m^2}~.
\end{equation} 
Remember that the second and third terms here originate from the Yukawa-type 
interaction (\ref{Vh}) and relativistic correction (\ref{Vr}) correspondingly.
To present the argument in the most clear form we neglected in Eq.(\ref{EanalytMin}) the 
potential, which stems from the gluon exchange. (The way to account it is discussed at the end of this subsection).

The value of the variational parameter $q$, which provides the minimum of $E_a(q)$ is
\begin{equation}
q_\text{min}\,=\, \frac{m}{9\alpha_h}\,\big( \,(\,1+9\alpha_h^2\,)^{1/2}-1\,\big)~.
\label{qmin}
\end{equation}
The corresponding value of the energy reads
\begin{equation}
\label{EanalytMin}
E_{a} \,=\,-\frac{m}9\,\Big(\,2\frac{ (\,1+9\alpha_h^2\,)^{3/2}-1}  {27\alpha_h^2}-1\,\Big)~.
\end{equation} 
Remember that $\alpha_h=m^2/(4\pi v^2)$ is a function of $m$. Hence, Eq.(\ref{EanalytMin})
provides the binding energy of the two-fermion state presenting it as a simple analytical function of $m$.
Its asymptotic behavior at small values of $\alpha_h<1$ is
\begin{align}
E_{a}\,\approx\,-\frac{m^5}{64\pi^2v^4}\Big(1-\frac{3m^4}{32\pi^2v^4}+\dots\Big)~.
\label{smallAlpha}
\end{align}
At large $\alpha_h>1$ we find
\begin{align}
E_{a}\,\approx\,-\frac{m^3}{18\pi v^2}\Big(1-\frac{2\pi v^2}{m^2}+\dots\Big)~.
\label{largeAlpha}
\end{align}
It is instructive to compare these results with the case when the relativistic correction, the last third term in Eq.(\ref{Eanalyt}),
is neglected. Then clearly the resulting two terms constitute the Coulomb-like problem. 
Their minimization gives the conventional Coulomb-type  energy
\begin{align}
E_{C}\,=\,-\frac{m^5}{64\pi^2v^4}~.
\label{coulomb}
\end{align}
Compare now Eqs.(\ref{EanalytMin}) and (\ref{coulomb}). 
The first takes into account the
repulsive relativistic correction, the second neglects it. 
Eq.(\ref{smallAlpha}) shows that at small $\alpha_h$ there is little
difference between the two cases. However, with the increase of the mass the 
distinction becomes prominent, compare (\ref{largeAlpha}) and (\ref{coulomb}).
Fig. \ref{fig8} illustrates this discrepancy. 
The dashed and solid lines there both represent
the variational energy, in which the relativistic correction is included.
The difference is that the solid line is taken for $m_h\rightarrow 0$, when 
analytical  Eq.(\ref{EanalytMin}) is valid, while the dashed one is for $m_h=100$ GeV, and was calculated numerically previously, see the data shown by the dashed line in Fig. \ref{fig3}.
The dotted line in Fig. \ref{fig8} shows the energy from Eq.(\ref{coulomb}), which neglects
the relativistic correction.

From Fig. \ref{fig8} we see again, this time basing the argument on clear analytical results, that the relativistic correction is important; it substantially reduces the binding energy of the two fermions. 
As a result the system remains in the nonrelativistic state for much larger masses, then one could have anticipated. To illustrate this point let us find the value of $m$, at which the
the energy in (\ref{EanalytMin}) equals the fermion mass. The condition
$|E_a|=m$ implies $\alpha_h\approx 4.97$, which corresponds to the huge mass
$m\approx 15$ TeV. For masses beyond this limit the 
relativistic physics takes over, and the developed in the present work approach, 
which is constructed on the nonrelativistic basis, is not applicable.
Remember though the more conservative estimate (\ref{VrT2}) for the mass, which limits the applicability of our results at $m\lesssim 0.8$ TeV. The just presented  arguments make it tempting to contemplate the possibility that the theory presented remains valid for  masses beyond the $\sim 1$ TeV limit, but in order to support this claim one should apply more sophisticated methods, possibly the Bethe-Salpeter equation, than the ones developed in the present work. 

Fig. \ref{fig8} shows that the dependence of the binding energy on $m_h$ is sizable, but still can be considered as a perturbations.
There is an interesting implication. Remember that previously it was found also that the gluon exchange, the radiative correction and the kinematic correction $-p^4/(4m^3)$ are all either very small, or at least not pronounced. Hence, we can treat them all as 
perturbations. In order to develop the first order perturbation theory we can
take $q_\text{min}$, which was found neglecting these perturbations, 
and substitute it into the complete expression for the variational energy, which takes 
all these corrections into consideration.

\begin{figure}[tbh]
  \centering \includegraphics[ height=5.2cm, keepaspectratio=true, angle=0]{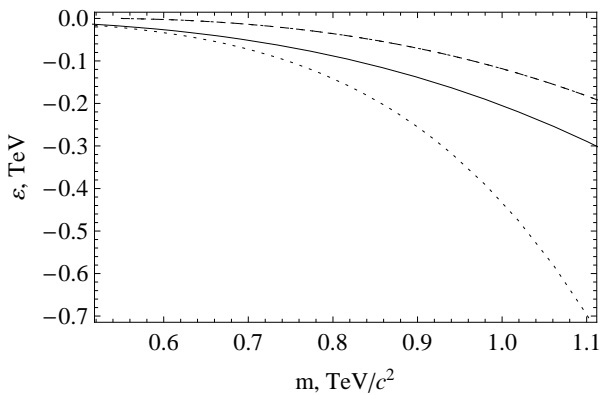}
\vspace{0cm}
\caption{
The role of relativistic correction.
Solid line - the variational energy from (\ref{EanalytMin}), which presumes  $m_h=0$ and accounts for the relativistic correction;
dashed line - the variational energy calculated from (\ref{minEq}) presuming $m_h=100$ GeV and taking the relativistic correction into account
(but without the radiative correction, same data is shown by the thin dashed line in Fig. \ref{fig3});
dotted line - the variational energy (\ref{coulomb}), in which $m_h=0$ and the relativistic correction is neglected.
}\label{fig8} 
   \end{figure}
   \noindent

Remembering that our variational approximation matches well the quantum mechanical calculations ($\le 20\%$), we can state that a reliable estimate for the energy $\varepsilon$ of the two heavy quarks bound via the Higgs boson exchange provides a simple, transparent analytical formula 
\begin{equation}
\varepsilon\approx E(q_\text{min})~, 
\label{analytical}
\end{equation}
in which $E(q)$ was found in (\ref{Eimproved}) while $q_\text{min}$ is given in (\ref{qmin}).
A more accurate value of $q_\text{min}$ which takes into account $m_h$ and the long-range 
(gluon) potential may be obtained using the results of section \ref{nonrel}:
\begin{equation}
q_\text{min}\,=\, \frac{m}{9 b}\,\big( \,(\,1+9 b^2\,)^{1/2}-1\,\big)~,
\label{qmin1}
\end{equation} 
where $b=(\tilde \alpha_h +a)$, see Eq. (\ref{qapprox}). For $m_h=a=0$ this result coincides with (\ref{qmin}).

\section{conclusion}

We discuss the binding of two heavy fermions due to the Higgs boson exchange.
Since the effective coupling constant increases with the fermion mass, 
the relativistic corrections are important. 
We find that they are positive and significantly reduce the binding energy. 
However, their influence on  the critical mass $m_{cr}$ for the fermion binding proves to 
be insignificant. 

The radiative corrections are found to play a minor role for all the interval of energies studied, at least up to $m=2$ TeV. In the case of heavy charged leptons the Higgs exchange dramatically increases the binding energy when mass $m$ approaches  $m_{cr}$. 
A very significant increase also happens for quarks when $m>m_{cr}$. The shift 
of energy, which is produced by the gluon exchange in the system bound mostly due to the Higgs exchange,  exceeds greatly the scale of energies typical to the systems bound due to the gluon  exchange alone.
 
The Higgs exchange also strongly increases the bound fermion density
 $\psi^2(0)$ and consequently the creation and annihilation widths for the bound state. The widths calculation will be performed in a separate publication. 
However, there is no fundamental reason, which would make impossible the detection of the bound states considered at LHC if the heavy fermions do exist. 
 
Note that we did not calculate the splitting of the spin $S=1$ and $S=0$ bound states. The relative value of this splitting $ \Delta E/2m$  should be significantly smaller then the leading relativistic corrections since it does not appear in the order $v^2/c^2$ in the Higgs exchange interaction, which is in difference with systems bound via the 
interactions produces by photon and gluon exchanges.
 
Note also that if there exist different heavy fermions they all contribute approximately equally to the polarization operator and vertex in Eqs.(\ref{P}), (\ref{fT}). This unusual property may be used to measure the number of heavy particles via the Higgs-dependent radiative corrections.

\section{acknowledgement}

This work is supported by the Australian Research Council.

\appendix

\section{Retardation and relativistic corrections} 

Taking Eq.(\ref{Int1}) for the relativistic and radiative correction
we apply the conventional for these problems technique replacing $\omega^2$ in the matrix element of the interaction 
by  the two commutators of the scalar  interaction vertex with the Hamiltonian
(compare e. g. \cite{QED} Section 83, \cite{QED1} Section 38; one can also compare the quantum approach with the classical description in \cite{QEDII} Section 65)
\begin{align}
\label{commutator} 
&\omega^2 \,
 \langle \,{\bf p}_1^\prime\,|\,\beta_1 \,e^{i{\bf q r}_1}|\,{\bf p}_1 \,\rangle
 \,\,
  \langle \,{\bf p}_2^\prime\,|\,\beta_2 \,e^{-i{\bf q r}_1}|\,{\bf p}_2 \,\rangle
\\
&
 \nonumber
 =-\langle \,{\bf p}_1^\prime\,|\,\,[H_1,\beta_1 \,e^{i{\bf q r}_1}]\,\,|\,{\bf p}_1 \,\rangle
  \,\,\langle \,{\bf p}_2^\prime\,|\,\,[H_2,\beta_2 \,e^{-i{\bf q r}_1}]\,\,|{\bf p}_2 \,\rangle~.
\end{align}   
Here $|{\bf p}_i^\prime\rangle$ and $|{\bf p}_i\rangle $, $i=1,2$ are the Dirac wave functions describing the scattering states of the two fermions, and $H_i\,=\, 
{\bm \alpha}_i \cdot\,{\bm  p}_i+\beta_i m$ are the Hamiltonians for these fermions.


Calculating the commutators in (\ref{commutator}) one finds
for the matrix element, which describes the scattering
\begin{align}
V({\bf p}_1,{\bf p}_2,{\bf q})
=&-{g_h^2}\Bigg[\frac{\beta_1 \beta_2}{{ {\bf q}^2+m_h^2}}
\label{ppq}
\\
&-
\frac{ 
\big(\bm{\gamma}_1\!\cdot\!(2{\bf p}_1+ {\bf q})\big) 
~\big(\bm{\gamma}_2\!\cdot\!(2{\bf p}_2- {\bf q} )\big) 
}
{ ({\bf q}^2+m_h^2)^2}
\Bigg]
\nonumber
\end{align}
In the center of mass reference frame ${\bf p}={\bf p}_1=-{\bf p}_1$ it is reduced
to $V({\bf q})$ in Eq.(\ref{V-4spinors}).
The potentials in (\ref{ppq}) and  (\ref{V-4spinors}) should be sandwiched
between the Dirac spinors describing the two fermions.
Fulfilling the Foldy-Wouthuysen transformation (\ref{uw}) we find that the matrix elements
of the beta-matrices in the nonrelativistic region can be replaced by the following expressions
\begin{align}
\label{abeta1}
&\beta_1\rightarrow 1-\frac{{\bf p}_1^2}{2m^2}-\frac{{\bf p}_1 {\bf \cdot\, q}}{2m^2}
-\frac{{\bf q}^2}{8m^2}-i\frac{({\bf q}\times{\bf p}_1) \cdot\,{\bm \sigma}_1}{4m^2}
\\
\label{abeta2}
&\beta_2\rightarrow 1-\frac{{\bf p}_2^2}{2m^2}+\frac{{\bf p}_2 {\bf \cdot\, q}}{2m^2}
-\frac{{\bf q}^2}{8m^2}+i\frac{({\bf q}\times{\bf p}_2) \cdot\,{\bm \sigma}_2}{4m^2}
\end{align}
In (\ref{beta-expansion}) these expressions are presented in the center of mass reference frame. Similarly we find that the matrix elements of gamma-matrices can be replaced as follows
\begin{align}
\label{agamma1}
&{\bm \gamma}_1\cdot (2{\bm p}_1+{\bf q})
\rightarrow -\frac{{\bf q}\cdot (2{\bm p}_1+{\bf q}) }{2m}~,
\\
\label{agamma2}
&{\bm \gamma}_2\cdot (2{\bm p}_2-{\bf q})
\rightarrow ~~\frac{{\bf q}\cdot (2{\bm p}_2-{\bf q}) }{2m}~.
\end{align}
Note the dependence of the matrix elements of gamma-matrices on ${\bf q}$.
Another notable feature is an absence of the spin dependence. Generically,
the spin variable is certainly 
present in the matrix elements of gamma-matrices
$
{\bm \gamma}_j\,\rightarrow -\big(\mp{\bf q}+i (2{\bm p}_j
\pm{\bf q})\cdot\, {\bm \sigma}_j\,\big)/(2m),
$
the upper and lower signs here are for $j=1$ and $2$ respectively.
However, the scalar products with the vectors $2{\bm p}_j\pm{\bf q}$, which
appear in (\ref{agamma1}) and (\ref{agamma2}), eradicate this dependence. 
This complies with the scalar nature of the Higgs boson.
It also explains why the simplified version of the discussion,
which is adopted in subsection \ref{Relativistic-retardation}
and does not appeal to the traditional commutator-based technique (\ref{commutator})
is able to match the results of a more robust
approach taken here.
The only available in the problem spin dependence arises from the beta-matrices in Eq.(\ref{abeta1}), (\ref{abeta2}) and eventually leads to the spin-orbit interaction in 
Eq.(\ref{L}).

We substitute Eqs.(\ref{abeta1})-(\ref{agamma2}) into (\ref{ppq}) deriving
\begin{align}
\nonumber
V({\bf p}_1,&{\bf p}_2,{\bf q})
=-{g_h^2}\Bigg[
\frac{1}
{ {\bf q}^2+m_h^2 }
\,\Big(\,1-\frac{ {\bm p}_1^2+{\bm p}_2^2 }{2m^2}
-\frac{ {\bm q}^2}{4m^2}
\\
\nonumber
&-\frac{ ({\bm p}_1-{\bm p}_2) \cdot {\bm q}}{2m^2}
-i\frac{ ({\bm q}\times {\bm p}_1)\cdot {\bm \sigma}_1-
({\bm q}\times {\bm p}_2)\cdot {\bm \sigma}_2}{4m^2}
 \,\Big)
 \\
&+
\frac{ 
\big(\,\bm {q}\!\cdot\!(2{\bf p}_1+ {\bf q})\,\big) 
~\big(\,\bm{q}\!\cdot\!(2{\bf p}_2- {\bf q} )\,\big) 
}
{ 4m^2\,({\bf q}^2+m_h^2)^2}
\Bigg]~.
\end{align}
After that
we adopt the center of mass reference frame and perform the Fourier transform
over ${\bf q}$ to the coordinate representation ${\bf r}$. Straightforward though lengthy calculations lead to Eqs.  (\ref{Vr}}), (\ref{D}), and (\ref{L}).


\end{document}